\documentclass[prl,aps,twocolumn,showpacs,superscriptaddress,floatfix]{revtex4-1}  
\usepackage{graphicx}\usepackage[dvips]{epsfig}
\usepackage{amssymb,amsmath,latexsym,bm}
\usepackage{newtxtext} 
\usepackage[shortcuts]{extdash}
\usepackage{colortbl}
\usepackage{multirow}
\usepackage{verbatim}
\usepackage{makecell}
\usepackage{pdfpages}

\makeatletter
\AtBeginDocument{\let\LS@rot\@undefined}
\makeatother

\renewcommand{\vec}[1]{\bm{\mathrm{#1}}}
\newcommand{\vhat}[1]{\hat{\bm{\mathrm{#1}}}}
\let\epsilon\varepsilon
\DeclareMathOperator{\sech}{sech}

\begin{document}
\title{Roles of chiral renormalization on magnetization dynamics in chiral magnets}
\author{Kyoung-Whan Kim}%
\email{kyokim@uni-mainz.de}
\affiliation{Institut f\"{u}r Physik, Johannes Gutenberg-Universit\"{a}t Mainz, Mainz 55128, Germany}%
\author{Hyun-Woo Lee}%
\email{hwl@postech.ac.kr}
\affiliation{Department of Physics, Pohang University of Science and Technology, Pohang 37673, Korea}%
\author{Kyung-Jin Lee}%
\affiliation{Department of Materials Science and Engineering, Korea
University, Seoul 02841, Korea}%
\affiliation{KU-KIST Graduate School of Converging Science and Technology,
Korea University, Seoul 02841, Korea}%
\author{Karin Everschor-Sitte}%
\affiliation{Institut f\"{u}r Physik, Johannes Gutenberg Universit\"{a}t Mainz, Mainz 55128, Germany}%
\author{Olena Gomonay}%
\affiliation{Institut f\"{u}r Physik, Johannes Gutenberg Universit\"{a}t Mainz, Mainz 55128, Germany}%
\affiliation{National Technical University of Ukraine ``KPI," Kyiv 03056, Ukraine}%
\author{Jairo Sinova}%
\affiliation{Institut f\"{u}r Physik, Johannes Gutenberg Universit\"{a}t Mainz, Mainz 55128, Germany}%
\affiliation{Institute of Physics, Academy of Sciences of the Czech Republic, Cukrovarnick\'{a} 10, 162 53 Praha 6, Czech Republic}
\date{\today}

\begin{abstract}
In metallic ferromagnets, the interaction between local magnetic moments and conduction electrons renormalizes parameters of the Landau-Lifshitz-Gilbert equation such as the gyromagnetic ratio and the Gilbert damping, and makes them dependent on the magnetic configurations. Although the effects of the renormalization for nonchiral ferromagnets are usually minor and hardly detectable, we show that the renormalization does play a crucial role for chiral magnets. Here the renormalization is chiral and as such we predict experimentally identifiable effects on the phenomenology of magnetization dynamics. In particular, our theory for the self-consistent magnetization dynamics of chiral magnets allows for a concise interpretation of domain wall creep motion. We also argue that the conventional creep theory of the domain wall motion, which assumes Markovian dynamics, needs critical reexamination since the gyromagnetic ratio makes the motion non-Markovian. The non-Markovian nature of the domain wall dynamics is experimentally checkable by the chirality of the renormalization.
\end{abstract}


\maketitle

Renormalization is a useful concept to understand interaction effects between a physical system and its environment. In metallic ferromagnets, magnetic moments experience such renormalization due to their coupling to conduction electrons through exchange interactions. Spin magnetohydrodynamic theory~\cite{Tserkovnyak09PRB,Wong09PRB,Zhang09PRL} examines the renormalization of dynamical parameters in the Landau-Lifshitz-Gilbert (LLG) equation as follows. Magnetization dynamics exerts a spin motive force (SMF)~\cite{Volovik87JPC,Barnes07PRL} on conduction electrons, and the resulting spin current generates spin-transfer torque (STT)~\cite{Slonczewski96JMMM,Berger96PRB,Ralph08JMMM} that affects the magnetization dynamics \emph{itself}. This self-feedback of magnetization dynamics~\cite{Lee13PR} renormalizes the Gilbert damping and the gyromagnetic ratio. However, its consequences rarely go beyond quantitative corrections in nonchiral systems~\cite{Kim11CAP,Kim11PRB,Kim15PRBa,Yuan16PRB,Cheng16JPD} and are commonly ignored.

Chiral magnets are ferromagnets that prefer a particular chirality of magnetic texture due to spin-orbit coupling (SOC) and broken inversion symmetry. Examples include ferromagnets in contact with heavy metals, such as Pt~\cite{Yang15PRL} and those with noncentrosymmetric crystal structures~\cite{Muhlbauer09S}. Magnetization dynamics in chiral magnets are usually described by generalizing the conventional LLG equation to include the chiral counterpart of the exchange interaction called the Dzyaloshinskii-Moriya interaction (DMI)~\cite{Dzyaloshinsky58PCS,Moriya60PR,Fert80PRL} and that of STT called spin-orbit torque (SOT)~\cite{Miron11N,Emori13NM,Liu12S,Ryu13NN}. This description is incomplete, however, since it ignores the renormalization by the self-feedback of magnetization dynamics. Although the renormalization in chiral magnets has been demonstrated theoretically for a few specific models~\cite{Kim12PRL,Akosa16PRB,Gungordu16PRB,Freimuth17PRB}, most experimental analyses of chiral magnets do not take into account the renormalization effect.

In this work, we demonstrate that the renormalization in chiral magnets should be chiral regardless of microscopic details and these effects should be nonnegligible in chiral magnets with large SOT observed in many experiments~\cite{Liu12S,Emori13NM,Ryu13NN,Kurebayashi14NN,Mellnik14N,Miron11NM}. Unlike in nonchiral systems, the chiral renormalization generates experimentally identifiable effects by altering the phenomenology of magnetization dynamics. This provides a useful tool to experimentally access underlying physics. We illustrate this with the field-driven magnetic domain wall (DW) motion with a controllable chirality by an external magnetic field~\cite{Thiaville12EPL,Je13PRB}. We find that not only is the steady state DW velocity chiral due to the chiral damping~\cite{Akosa16PRB}, but also the effective mass of the DW~\cite{Doring48NA} is chiral due to the chiral gyromagnetic ratio. The chiral gyromagnetic ratio also significantly affects the DW creep motion, which is one of the techniques to measure the strength of the DMI~\cite{Je13PRB}. We argue that the chiral gyromagnetic ratio is the main mechanism for the non-energetic chiral DW creep velocity~\cite{Jue15NM}, contrary to the previous attribution to the chiral damping~\cite{Jue15NM,Akosa16PRB}.
We also highlight the importance of the tilting angle excitation and its delayed feedback to the DW motion. This has been ignored in the traditional creep theory~\cite{Lemerle98PRL,Chauve00PRB} for a long time, since its effects merely alter the velocity prefactor which is indistinguishable from other contributions, such as the impurity correlation length~\cite{Gorchon14PRL}. However, in chiral magnets, it is distinguishable by measuring the DW velocity as a \emph{function} of chirality (not a single value). 

\begin{figure}
	\includegraphics[width=8.6cm]{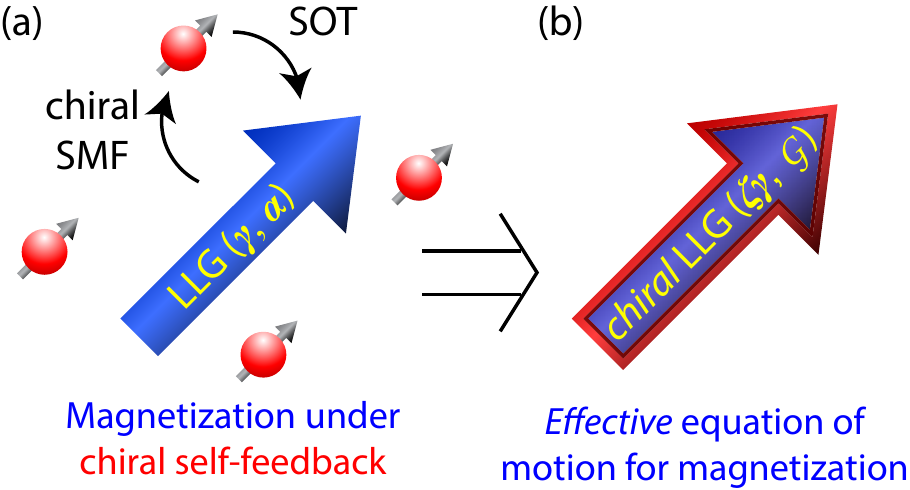}
	\caption{(a) Magnetization dynamics described by the unrenormalized LLG equation. The dynamics of magnetization and that of electrons are coupled to each other by the exchange interaction. (b) After tracing out the electron degree of freedom, the gyromagnetic ratio ($\zeta\gamma$) and the magnetic damping ($\mathcal{G}$) are chirally renormalized [Eq.~(\ref{Eq:renormalized Landau-Lifshitz-Gilbert})].
	}
	\label{Fig:concept}
\end{figure}

To get deep insight into the chiral renormalization, we adopt the self-feedback mechanism of magnetization dynamics through conduction electrons and develop a general, concise, and unified theory for chiral magnets. There are several previous reports on the anisotropic or chiral renormalization of the magnetic damping~\cite{Kim12PRL,Tatara13PRB,Gungordu16PRB,Akosa16PRB} and the gyromagnetic ratio~\cite{Bijl12PRB,Freimuth17PRB,Tatara13PRB} in the Rashba model~\cite{Bychkov84JETP}. To unify and generalize the previous works, we start from the general Onsager reciprocity relation and predict all the core results of the previous reports. Our theory can be generalized to situations with any phenomenological spin torque expression, which can even be determined by symmetry analysis and experiments without knowing its microscopic mechanism. We provide a tabular picture (See Table~\ref{Tab:feedback} below) for physical understanding of each contribution to the chiral renormalization. Furthermore, one can utilize the generality of the Onsager relation to include magnon excitations~\cite{Gungordu16PRB}, thermal spin torques~\cite{Hatami07PRL}, and even mechanical vibrations~\cite{Matsuo11PRL} in our theory.

To examine the consequences of the chiral renormalization, we start from the following renormalized LLG equation, which we derive in the later part of this paper,
\begin{equation}
(\zeta\gamma)^{-1}\cdot\partial_t\vec{m}=-\vec{m}\times\vec{H}_{\rm eff}+\gamma^{-1}\vec{m}\times\mathcal{G}\cdot\partial_t\vec{m}+\gamma^{-1}\vec{T}_{\rm ext},\label{Eq:renormalized Landau-Lifshitz-Gilbert}
\end{equation}
where $\vec{m}$ is the unit vector along magnetization, $\gamma$ is the unrenormalized gyromagnetic ratio, $\vec{H}_{\rm eff}$ is the effective magnetic field, and $\vec{T}_{\rm ext}$ refers to spin torque induced by an external current. $\zeta$ and $\mathcal{G}$, which are generally tensors and functions of $\vec{m}$ and its gradients, address respectively the renormalization of the gyromagnetic ratio and the magnetic damping, depicted in Fig.~\ref{Fig:concept}. If the renormalization is neglected, Eq.~(\ref{Eq:renormalized Landau-Lifshitz-Gilbert}) reduces to the conventional LLG equation with $\zeta=1$ and $\mathcal{G}=\alpha$, where $\alpha$ is the unrenormalized Gilbert damping. Otherwise $\zeta$ and $\mathcal{G}$ are dependent on the chirality of magnetic texture. At the end of this paper, we show that the chiral renormalization is completely fixed once the expressions of STT and SOT are given.

\begin{figure}
	\includegraphics[width=8.6cm]{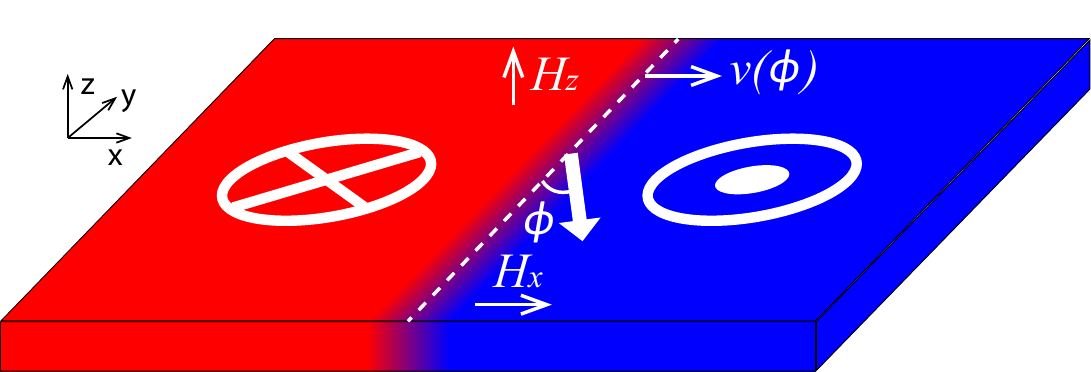}
	\caption{Chiral dynamics of a DW between domains with $\vec{m}=\mp\vhat{z}$ (red and blue respectively). The DW chirality is characterized by the DW tilting angle $\phi$ [the positivity (negativity) of $\phi$ corresponds to the left-handed (right-handed) chirality], and can be controlled by an in-plane field ($H_x$). The DW motion is driven by an applied field ($H_z$). Measuring the DW velocity as a function of $\phi$ (or $H_x$), the difference between $v(\phi)$ and $v(-\phi)$ gives the information of the chiral renormalization.
	}
	\label{Fig:DW}
\end{figure}

We first examine implications of the chiral renormalization on a few exemplary types of field-driven DW dynamics (Fig.~\ref{Fig:DW}).
We start from $\vec{H}_{\rm eff}=\vec{H}_0+\vec{H}_{\rm ext}+\vec{H}_{\rm th}$, where $\vec{H}_0$ is the energetic contribution (without an external field), $\vec{H}_{\rm ext}=(H_x, 0, H_z)$ is the external field, and $\vec{H}_{\rm th}$ is a thermal fluctuation field. We use the DW profile $\vec{m}(x)=(\sin\phi\sech[(x-X)/\lambda], \cos\phi\sech[(x-X)/\lambda], \tanh[(x-X)/\lambda])$ where $X$, $\phi$ and $\lambda$ are the position, the tilting angle, and the width of the DW, respectively. Taking $X$ and $\phi$ as the collective coordinates, Eq.~(\ref{Eq:renormalized Landau-Lifshitz-Gilbert}) gives
\begin{subequations}
	\label{Eq:Thiele}
	\begin{align}
	\frac{\alpha_{\rm eff}^X}{\lambda}\frac{dX}{dt}+\frac{1}{\zeta_{\rm eff}}\frac{d\phi}{dt}&=\mathcal{F}_X+\xi_X,\label{Eq:Thiele1}\\
	-\frac{1}{\zeta_{\rm eff}}\frac{dX}{dt}+\alpha_{\rm eff}^\phi\lambda\frac{d\phi}{dt}&=\mathcal{F}_\phi+\xi_\phi,\label{Eq:Thiele2}
	\end{align}
\end{subequations}
where $\mathcal{F}_{X/\phi}=(\gamma/2)\int (\vec{H}_0+\vec{H}_{\rm ext})\cdot(\partial_{X/\phi}\vec{m})dx$ refer to the force on $X$ and $\phi$. $\xi_{X/\phi}=(\gamma/2)\int\vec{H}_{\rm th}\cdot(\partial_{X/\phi}\vec{m})dx$ is the thermal force on $X$ and $\phi$.

\begin{figure}
	\includegraphics[width=8.6cm]{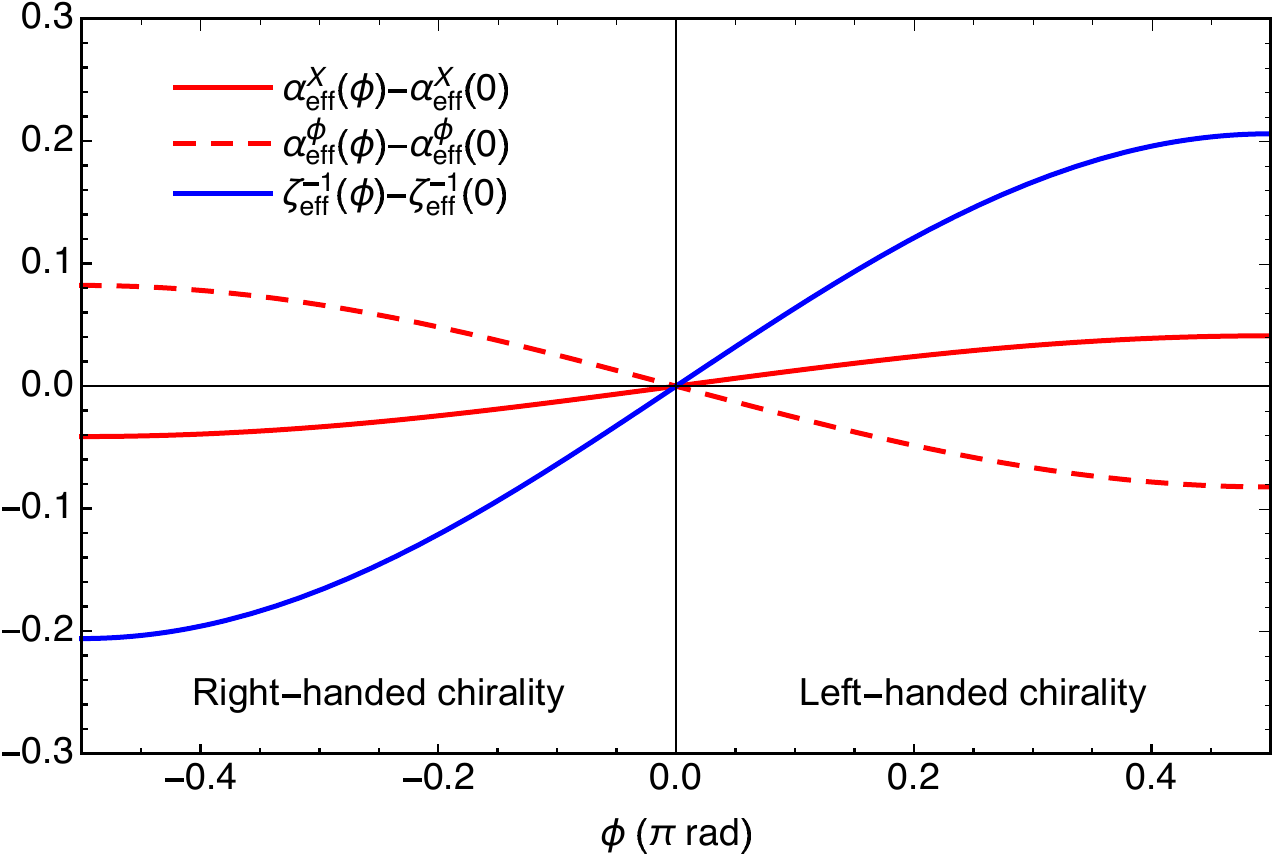}
	\caption{The effective dynamical parameters, $\alpha_{\rm eff}^X$ (the red, solid curve), $\alpha_{\rm eff}^\phi$ (the red, dashed curve), and $\zeta_{\rm eff}^{-1}$ (the blue curve), as a function of the DW tilting angle $\phi$.
	We take the phenomenological expression of spin torque in magnetic bilayers~\cite{Liu12S,Emori13NM,Ryu13NN,Kurebayashi14NN}, which is a typical example with large SOT: $\vec{T}=(\gamma\hbar/2eM_s)\{(\vec{j}_s\cdot\nabla)\vec{m}-\beta_1\vec{m}\times(\vec{j}_s\cdot\nabla)\vec{m}+k_{\rm SO}(\vhat{z}\times\vec{j}_s)\times\vec{m}-\beta_2k_{\rm SO}\vec{m}\times[(\vhat{z}\times\vec{j}_s)\times\vec{m}]\}$, where each term refers to the adiabatic STT~\cite{Tatara04PRL}, nonadiabatic STT~\cite{Zhang04PRL,Kim15PRB}, fieldlike SOT~\cite{Manchon08PRB,Matos-A09PRB}, and dampinglike SOT~\cite{Wang12PRL,Kim12PRB,Pesin12PRB,Kurebayashi14NN}, induced by the spin current $\vec{j}_s$. Here, $M_s=1000~\mathrm{emu/cm^3}$ is the saturation magnetization, $e>0$ is the (negative) electron charge, $\vhat{z}$ is the interface normal direction, $k_{\rm SO}=1.3~(\mathrm{nm})^{-1}$ characterizes the strength of the SOT.
	We take $\beta_1=0.05$, $\beta_2=5$, $\lambda=8~\mathrm{nm}$, and the electrical conductivity $\sigma_0^{-1}=6~\mathrm{\mu\Omega cm}$. The parameters are on the order of the typical values for Pt/Co systems~\cite{Kim13PRL,Zhang04PRL,Miron11NM}.
	}
	\label{Fig:plot}
\end{figure}

The effective damping $\alpha_{\rm eff}^{X/\phi} $and the gyromagnetic ratio $\zeta_{\rm eff}$ are given by
\begin{subequations}
	\label{Eq:effective parameters}
	\begin{align}
	\alpha_{\rm eff}^X&=\frac{\lambda}{2}\int\left( \partial_X\vec{m}\cdot\mathcal{G}\cdot\partial_X\vec{m}\right)dx,\allowdisplaybreaks\\
	\alpha_{\rm eff}^\phi&=\frac{1}{2\lambda}\int\left( \partial_\phi\vec{m}\cdot\mathcal{G}\cdot\partial_\phi\vec{m}\right)dx,\allowdisplaybreaks\\
	\zeta_{\rm eff}^{-1}&=\frac{1}{2}\int\left[ (\vec{m}\times\partial_{\phi}\vec{m})\cdot\zeta^{-1}\cdot\partial_X\vec{m}\right]dx.
	\end{align}
\end{subequations}
Note that without the chiral renormalization, Eq.~(\ref{Eq:Thiele}) reduces to the Thiele equations~\cite{Thiele73PRL} with $\alpha_{\rm eff}^{X/\phi}=\alpha$ and $\zeta_{\rm eff}=1$. We emphasize that $\alpha_{\rm eff}^{X/\phi}$ and $\zeta_{\rm eff}$ depend on the tilting angle $\phi$ and thus on the chirality of the DW. Figure~\ref{Fig:plot} shows the $\phi$ dependencies of these parameters. The asymmetric dependences on $\phi$ confirm their chiral dependences. 
Note that, even for purely field-driven DW motion, the chiral dependences of the parameters are determined by the expression of \emph{current-induced} spin torque.

We first consider the steady-state dynamics of DW in the flow regime, where the effects of the pinning and the thermal forces are negligible. Then, translational symmetry along $X$ guarantees the absence of contribution from $\vec{H}_0$ to $\mathcal{F}_X$, thus only the external field contribution survives in the right-hand side of Eq.~(\ref{Eq:Thiele1}), $\mathcal{F}_X+\xi_X\approx-\gamma H_z$. In a steady state ($d\phi/dt=0$), Eq.~(\ref{Eq:Thiele1}) gives the DW velocity as
\begin{equation}
v_{\rm flow}=-\frac{\gamma\lambda}{\alpha_{\rm eff}^X}H_z,\label{Eq:vDW flow}
\end{equation}
which is inversely proportional to the chiral damping $\alpha_{\rm eff}^X$ evaluated at the steady-state tilting angle $\phi_{\rm eq}$ for which $d\phi/dt=0$. As $\phi_{\rm eq}$ can be modulated by $H_x$, the measurement of $v_{\rm flow}$ as a function of $H_x$ provides a direct test of the chiral dependence on $\alpha_{\rm eff}^X$.

As an experimental method to probe the chiral dependence of $\zeta_{\rm eff}$, we propose the measurement of the DW mass, called the D\"{o}ring mass~\cite{Doring48NA}. It can be performed by examining the response of DW under a potential trap to an oscillating field $H_z$~\cite{Saitoh04N}. Unlike $v_{\rm flow}$, $\phi$ is not stationary for this case, and dynamics of it is coupled to that of $X$. Such coupled dynamics of $\phi$ and $X$ makes $\zeta_{\rm eff}$ relevant. In the Supplemental Material~\cite{Supple}, we integrate out the coupled equations [Eq.~(\ref{Eq:Thiele})] to obtain the effective D\"{o}ring mass,
\begin{equation}
m_{\rm DW}=\frac{1}{\zeta_{\rm eff}^2}\frac{2M_sS}{\gamma|\mathcal{F}_\phi'(\phi_{\rm eq})|},\label{Eq:chiral mass}
\end{equation}
where $S$ is the cross-sectional area of the DW. Here, $\zeta_{\rm eff}$ represents a measurement of its value for $\phi=\phi_{\rm eq}$, which can be varied by $H_x$. $m_{\rm DW}$ provides an experimental way to measure the chiral dependence of $\zeta_{\rm eff}$.

In the creep regime of the DW where the driving field is much weaker than the DW pinning effects, the implication of the chiral renormalization go beyond merely chiral corrections to the DW velocity. The recent controversies on the chiral DW creep speed $v_{\rm creep}$ measured from various experiments~\cite{Je13PRB,Jue15NM,Lavrijsen15PRB,Balk17PRL} require more theoretical examinations. Typically, $v_{\rm creep}$ is believed to follow the Arrhenius-like law $v_{\rm creep} = v_0 \exp(-\kappa H_z^{-\mu}/k_B T)$~\cite{Lemerle98PRL,Chauve00PRB}, where $v_0$ is a prefactor, $\mu$ is the creep exponent typically chosen to be 1/4~\cite{Cayssol04PRL}, and $\kappa$ is a parameter proportional to the DW energy density. Based on the observation that the DMI affects $\kappa$, an experiment~\cite{Je13PRB} attributed the chiral dependence of $v_{\rm creep}$ to the DMI. However later experiments~\cite{Jue15NM,Lavrijsen15PRB,Balk17PRL} found features that cannot be explained by the DMI. In particular, Ref.~\cite{Jue15NM} claimed that the chiral dependence of $v_{\rm creep}$ is an indication of the chiral damping~\cite{Akosa16PRB}, based on the observation $v_0 \propto (\alpha_{\rm eff}^X)^{-1}$. On the other hand, our analysis shows that the explanation of the chirality dependence may demand more fundamental change to the creep law, which assumes the dynamics of $\phi$ to be essentially decoupled from that of $X$ and thus irrelevant for $v_{\rm creep}$.
As a previous experiment on the DW creep motion in a diluted semiconductor~\cite{Yamanouchi07S} argued the coupled dynamics of $\phi$ and $X$ to be important, it is not \emph{a priori} clear whether the assumption of decoupling $X$ and $\phi$ holds in the creep regime.

We consider the coupling between the dynamics of $X$ and $\phi$ as follows. After the dynamics of $X$ excites $\phi$, the dynamics of $\phi$ results in a feedback to $X$ with a delay time $\tau$. Since the dynamics at a time $t$ is affected by its velocity at past $t-\tau$, it is non-Markovian. The traditional creep theory takes the Markovian limit ($\tau\to0$), thus $\phi=\phi_{\rm eq}$ at any instantaneous time, decoupled from the dynamics of $X$.
To show the crucial role of a finite feedback time $\tau$, we calculate the escape rate of the DW over a barrier, which is known to be proportional to $v_0$~\cite{Gorchon14PRL} and apply the Kramer's theory~\cite{Kramer40P} for barrier escape and its variations for non-Markovian systems~\cite{Grote80JCP,Pollak89JCP}. After some algebra in the Supplemental Material~\cite{Supple}, Eq.~(\ref{Eq:Thiele}) gives
\begin{equation}
v_0
\propto\left\{\begin{array}{lll}
\displaystyle(\alpha_{\rm eff}^X)^{-1}& \tau\nu_0\ll\zeta_{\rm eff}^2\alpha_{\rm eff}^X\alpha_{\rm eff}^\phi&(\text{Markovian}),\\
\displaystyle\zeta_{\rm eff}&\tau\nu_0\gtrsim\zeta_{\rm eff}^2\alpha_{\rm eff}^X\alpha_{\rm eff}^\phi&(\text{non-Markovian}),
\end{array}\right.
\label{Eq:DW creep velocity}
\end{equation}
where $\nu_0$ is called the reactive frequency~\cite{Pollak89JCP} and is on the order of $2\pi$ times the attempt frequency ($\approx1~\mathrm{GHz}$~\cite{Gorchon14PRL}). We emphasize that the two regimes show very different behavior in the sense of underlying physics as well as phenomenology. The validity of the Markovian assumption depends on the time scale of $\tau$ compared to $\zeta_{\rm eff}^2\alpha_{\rm eff}^X\alpha_{\rm eff}^\phi$. Since the damping is small, it is not guaranteed for our situation to be in the Markovian regime. Indeed, we demonstrate in the Supplemental Material~\cite{Supple} that the second regime (non-Markovian) in Eq.~(\ref{Eq:DW creep velocity}) is more relevant with realistic values, thus the chirality of $v_0$ mainly originates from the gyromagnetic ratio, not the damping~\cite{Jue15NM}. One can measure the chiral dependence of $\alpha_{\rm eff}^X$ and $\zeta_{\rm eff}$ from the flow regime [Eqs.~(\ref{Eq:vDW flow}) and (\ref{Eq:chiral mass})] and compare their chiral dependences to the creep regime to observe the non-Markovian nature of the DW dynamics. This advantage originates from the possibility that one can measure the DW velocity as a \emph{function} of chirality, in contrast to nonchiral magnets where one measures the DW velocity as a single value.

So far, we present the role of the chiral renormalization for given renormalized tensors $\mathcal{G}$ and $\zeta$. To provide underlying physical insight into it, we present a analytic derivation of Eq.~(\ref{Eq:renormalized Landau-Lifshitz-Gilbert}) in general situations. We start from the LLG equation $\gamma^{-1}\partial_t\vec{m}=-\vec{m}\times\vec{H}_{\rm eff}+\gamma^{-1}\alpha\vec{m}\times\partial_t\vec{m}+\gamma^{-1}\vec{T}$ and refer to the scenario illustrated in Fig.~\ref{Fig:concept}. Note that $\vec{T}$ here includes a contribution from an internally generated SMF ($\vec{T}_{\rm int}$) as well as that from an external current [$\vec{T}_{\rm ext}$ in Eq.~(\ref{Eq:renormalized Landau-Lifshitz-Gilbert})]. We write down the spin torque in a general current-linear form $\vec{T}=-(\gamma\hbar/2eM_s)\vec{m}\times\sum_u\vec{A}_u(\vec{m})j_{s,u}$, where $u$ runs over $x,y,z$. Here the spin current $\vec{j}_s$ is split into an internally generated SMF~\cite{Volovik87JPC,Barnes07PRL} $\vec{j}_{s, \rm int}$ and the external current $\vec{j}_{s, \rm ext}$. The former is proportional to $\partial_t\vec{m}$, thus it renormalizes the gyromagnetic ratio and the damping. The latter generates $\vec{T}_{\rm ext}$ in Eq.~(\ref{Eq:renormalized Landau-Lifshitz-Gilbert}). The expression of $\vec{j}_{s, \rm int}$ is given by the Onsager reciprocity of STT and SMF~\cite{Tserkovnyak08PRB}: $j_{s,\mathrm{int}, u}=-(\sigma_0\hbar/2e)\vec{A}_u(-\vec{m})\cdot\partial_t\vec{m}$, where $\sigma_0$ is the charge conductivity~\cite{footnote}. Substituting this to $\vec{T}_{\rm int}=(\gamma\hbar/2eM_s)\vec{m}\times\sum_u\vec{A}_u(\vec{m})j_{s,\mathrm{int},u}$ gives the effective LLG equation $\gamma^{-1}\partial_t\vec{m}=-\vec{m}\times\vec{H}_{\rm eff}+\gamma^{-1}\vec{m}\times\mathcal{A}\cdot\partial_t\vec{m}+\gamma^{-1}\vec{T}_{\rm ext}$, where $\mathcal{A}=\alpha+\eta\sum_u \vec{A}_u(\vec{m})\otimes\vec{A}_u(-\vec{m})$, $\eta=\gamma\hbar^2\sigma_0/4e^2M_s$ and $\otimes$ is the direct tensor product. As a result, $\vec{T}_{\rm int}$ is taken care of by renormalizing $\alpha$ into $\mathcal{A}$ in the LLG equation.

The renormalized damping and gyromagnetic ratio are given by separating different contributions of $\mathcal{A}$ with different time reversal properties. A damping contribution is required to be dissipative (odd in time reversal), whereas a gyromagnetic term should be reactive (even in time reversal). Separating these gives Eq.~(\ref{Eq:renormalized Landau-Lifshitz-Gilbert}) where $\mathcal{G}=(\mathcal{A}+\mathcal{A}^T)/2$ and $\zeta^{-1}=1-\vec{m}\times(\mathcal{A}-\mathcal{A}^T)/2$. The particular choice for the adiabatic STT and the nonadiabatic STT  $\vec{A}_u(\vec{m})=\vec{m}\times\partial_u\vec{m}+\beta\partial_u\vec{m}$ reproduces the renormalized LLG equation for nonchiral systems~\cite{Tserkovnyak09PRB,Zhang09PRL,Wong09PRB}. When one uses $\vec{A}_u(\vec{m})$ for a particular chiral system, Eq.~(\ref{Eq:renormalized Landau-Lifshitz-Gilbert}) produces the effective LLG equation for it, as reported by a numerical study for a one-dimensional Rashba model~\cite{Freimuth17PRB}.

\begin{table}[b]
	\begin{tabular}{|c|c|c|c|c|}
		\hline&\multicolumn{4}{c|}{STT: $\vec{A}_x(\vec{m})$}\\\cline{2-5}
		\makecell{SMF:\\$\vec{A}_x(-\vec{m})$}&\makecell{Adiabatic\\$\vec{m}\times\partial_x\vec{m}$}&\makecell{\scalebox{1}[1.0]{Nonadiabatic}\\$\beta_1\partial_x\vec{m}$}&\makecell{FLT\\\scalebox{0.7}[1.0]{$k_{\rm SO}\vec{m}\times(\vhat{y}\times\vec{m})$}}&\makecell{DLT\\\scalebox{1}[1.0]{$\beta_2k_{\rm SO}\vhat{y}\times\vec{m}$}}\\\hline
		$\vec{m}\times\partial_x\vec{m}$&$\mathcal{G}$&$\zeta^{-1}$&\cellcolor[gray]{0.9}$\mathcal{G}$&\cellcolor[gray]{0.9}$\zeta^{-1}$\\\hline
		$-\beta_1\partial_x\vec{m}$&$\zeta^{-1}$&$\mathcal{G}$&\cellcolor[gray]{0.9}$\zeta^{-1}$&\cellcolor[gray]{0.9}$\mathcal{G}$\\\hline
		\scalebox{0.8}[1]{$k_{\rm SO}\vec{m}\times(\vhat{y}\times\vec{m})$}&\cellcolor[gray]{0.9}$\mathcal{G}$&\cellcolor[gray]{0.9}$\zeta^{-1}$&\cellcolor[gray]{0.6}$\mathcal{G}$&\cellcolor[gray]{0.6}$\zeta^{-1}$\\\hline
		\scalebox{1}[1.0]{$-\beta_2k_{\rm SO}\vhat{y}\times\vec{m}$}&\cellcolor[gray]{0.9}$\zeta^{-1}$&\cellcolor[gray]{0.9}$\mathcal{G}$&\cellcolor[gray]{0.6}$\zeta^{-1}$&\cellcolor[gray]{0.6}$\mathcal{G}$\\\hline
	\end{tabular}
	\caption{\label{Tab:feedback} Example characterization of contributions in $\vec{A}_x(\vec{m})\otimes\vec{A}_x(-\vec{m})$. Counting orders of gradients and $\vec{m}$ gives the conventional (white), chiral (lighter gray), or anisotropic (darker gray) contributions to the gyromagnetic ratio ($\zeta^{-1}$) or the damping ($\mathcal{G}$). The form of the fieldlike SOT (FLT) and dampinglike SOT (DLT) are taken from magnetic bilayers~\cite{Wang12PRL,Kim12PRB,Pesin12PRB,Kurebayashi14NN,Manchon08PRB,Matos-A09PRB} for illustration, but the characterization procedure works generally.}
\end{table}

In chiral magnets, it is known that spin torque includes two more contributions called fieldlike SOT~\cite{Manchon08PRB,Matos-A09PRB} and  dampinglike SOT~\cite{Wang12PRL,Kim12PRB,Pesin12PRB,Kurebayashi14NN}. The characterization of fieldlike and dampinglike SOT is regardless of the choice of SOC, since it is determined by the time reversal characteristic. Since $\vec{A}_u(\vec{m})$ consists of four contributions, there are 16 contributions in the feedback tensor $\Delta\mathcal{A}=\eta\sum_u\vec{A}_u(\vec{m})\otimes\vec{A}_u(-\vec{m})$ for each $u$. We tabulate all terms of $\Delta\mathcal{A}$ in Table~\ref{Tab:feedback}. The contributions with the white background are zeroth order in SOC but second order in gradient and are the conventional nonlocal contributions~\cite{Wang15PRB,Zhang09PRL}. Those with the lighter gray background are first order in gradient and chiral~\cite{Freimuth17PRB}. Those with the darker gray color are zeroth order in gradient and anisotropic~\cite{Hals14PRB}. In this way, our theory provides a unified picture on the previous works. Whether a term contributes to $\zeta^{-1}$ or $\mathcal{G}$ is determined by the order in $\vec{m}$. After a direct product of STT and SMF, a term even (odd) in $\vec{m}$ gives $\mathcal{G}$ ($\zeta^{-1}$), since it gives a time irreversible (reversible) contribution appearing in the LLG equation as $\vec{m}\times\mathcal{A}\cdot\partial_t\vec{m}$. The same analysis with simple order countings works for any $\vec{A}_u(\vec{m})$. It holds even if our theory is generalized to other physics, such as magnons~\cite{Gungordu16PRB}, thermal effects~\cite{Hatami07PRL}, and mechanical effects~\cite{Matsuo11PRL}.

As an example of applications of Table~\ref{Tab:feedback}, we adopt the spin-Hall-effect driven SOT~\cite{Seo12APL,Sinova15RMP,Liu12S}, where a large dampinglike SOT arises. From Table~\ref{Tab:feedback}, one can immediately figure out that the combination of the dampinglike SOT and the conventional SMF (the most top right cell) gives a chiral gyromagnetic ratio contribution. As another example, one notices that the combination of the dampinglike SOT and its Onsager counterpart (the fourth term in the SMF) gives an anisotropic damping contribution. Note that the Onsager counter part of the spin-Hall-effect driven SOT is the inverse spin Hall effect driven by spin pumping current generated by the magnetization dynamics. In this way, Table~\ref{Tab:feedback} provides useful insight for each contribution.

Table~\ref{Tab:feedback} also allows for making the general conclusion that the magnitude of the chiral gyromagnetic ratio is determined by the size of the dampinglike SOT ($\beta_2$) and that of the nonadiabatic STT ($\beta_1$). This is an important observation since many experiments on magnetic bilayers and topological insulators~\cite{Liu12S,Emori13NM,Ryu13NN,Kurebayashi14NN} shows a large dampinglike SOT. This conclusion is regardless of the microscopic details of the SOT, because a dampinglike contribution is solely determined by its time-reversal property.

To summarize, we demonstrate that the chiralities of the gyromagnetic ratio and Gilbert damping have significant implications which go further beyond merely the change in magnetization dynamics. The chirality plays an important role in investigating underlying physics because physical quantities, which were formerly treated as constants, can now be controlled through their chiral dependence. An example is the non-Markovian character of the DW creep motion, which is difficult to be verified in nonchiral systems. From the non-Markovian nature of the DW creep motion, we conclude that the non-energetic origin of chiral DW creep originates from the chiral gyromagnetic ratio rather than the chiral damping. We also provide a general, concise, and unified theory of their chiralities, which provide useful insight on the self-feedback of magnetization.

\begin{acknowledgments}
We acknowledge M.~D.~Stiles, Y.~Tserkovnyak, A.~Thiaville, S.\=/W.~Lee, V.~Amin, and D.\=/S.~Han for fruitful discussion. This work is supported by the Alexander von Humboldt Foundation, the ERC Synergy Grant SC2 (No. 610115), the Transregional Collaborative Research Center (SFB/TRR) 173 SPIN+X, and the German Research Foundation (DFG) (No. EV 196/2-1 and No. SI 1720/2-1). K.W.K also acknowledges support by Basic Science Research Program through the National Research Foundation of Korea (NRF) funded by the Ministry of Education (2016R1A6A3A03008831). H.W.L. was supported by NRF (2011-0030046). K.J.L was supported by NRF (2015M3D1A1070465, 2017R1A2B2006119).
\end{acknowledgments}



\section*{Supplementary material}

# Supplementary Materials for
## "Roles of chiral renormalization of magnetization dynamics in chiral magnets"

Kyoung-Whan Kim,[1] Hyun-Woo Lee,[2] Kyung-Jin Lee,[3,4] Karin Everschor-Sitte,[1] Olena Gomonay,[1,5] and Jairo Sinova[1,6]

[1]*Institut für Physik, Johannes Gutenberg Universität Mainz, Mainz 55128, Germany*
[2]*PCTP and Department of Physics, Pohang University of Science and Technology, Pohang 37673, Korea*
[3]*Department of Materials Science and Engineering, Korea University, Seoul 02841, Korea*
[4]*KU-KIST Graduate School of Converging Science and Technology, Korea University, Seoul 02841, Korea*
[5]*National Technical University of Ukraine "KPI", Kyiv 03056, Ukraine*
[6]*Institute of Physics, Academy of Sciences of the Czech Republic, Cukrovarnická 10, 162 53 Praha 6 Czech Republic*

## I. THE NON-MARKOVIAN NATURE OF THE DW DYNAMICS

### A. Integrating out $\phi$

In the linear response regime, we may take $\mathcal{F}_\phi \approx -|\mathcal{F}'_\phi(\phi_\mathrm{eq})|(\phi - \phi_\mathrm{eq})$ and the dynamical coefficients $\zeta_\mathrm{eff}$ and $\alpha_\mathrm{eff}^{X/\phi}$ to be evaluated at $\phi = \phi_\mathrm{eq}$. Without loss of generality, we assume the initial condition $X(0) = 0$ and $\phi(0) = \phi_\mathrm{eq}$. We then define the Laplace transforms $\mathcal{L}[f(t)](s) = \int_0^\infty e^{-st} f(t) dt$. We denote $\mathcal{L}[X] = Q$ and $\mathcal{L}[\phi - \phi_\mathrm{eq}] = P$. Then the Laplace transform of Eq. (2) in the main text is

$$\frac{s\alpha_\mathrm{eff}^X}{\lambda} Q + \frac{s}{\zeta_\mathrm{eff}} P = \mathcal{L}[\mathcal{F}_X] + \mathcal{L}[\xi_X], \tag{S1a}$$

$$-\frac{s}{\zeta_\mathrm{eff}} Q + s\alpha_\mathrm{eff}^\phi \lambda P = -|\mathcal{F}'_\phi(\phi_\mathrm{eq})| P + \mathcal{L}[\xi_\phi], \tag{S1b}$$

Eliminating $P$ in Eq. (S1) gives

$$\frac{1}{\zeta_\mathrm{eff}^2} \frac{s^2}{|\mathcal{F}'_\phi(\phi_\mathrm{eq})| + s\alpha_\mathrm{eff}^\phi \lambda} Q + \frac{s\alpha_\mathrm{eff}^X}{\lambda} Q = -\frac{\gamma H_z}{s} + \mathcal{L}[\mathcal{F}_\mathrm{pin}] + \left( \mathcal{L}[\xi_X] - \frac{s}{\zeta_\mathrm{eff}} \frac{\mathcal{L}[\xi_\phi]}{b + s\alpha_\mathrm{eff}^\phi \lambda} \right), \tag{S2}$$

which is an equation of $X$ only. Taking the inverse Laplace transform, we obtain the following non-Markovian equation:

$$\frac{1}{\lambda} \int_0^t f(t-u) X'(u) du = \mathcal{F}_X + \tilde{\xi}_X(t). \tag{S3}$$

Here $f(t)$ is a feedback function from $\phi$, whose explicit form is

$$f(t) = \mathcal{L}^{-1}\left[ \alpha_\mathrm{eff}^X + \frac{1}{\zeta_\mathrm{eff}^2 \alpha_\mathrm{eff}^\phi} \frac{s\tau}{1 + s\tau} \right] = \left( \alpha_\mathrm{eff}^X + \frac{1}{\zeta_\mathrm{eff}^2 \alpha_\mathrm{eff}^\phi} \right) \delta(t) - \frac{1}{\zeta_\mathrm{eff}^2 \alpha_\mathrm{eff}^\phi \tau} e^{-t/\tau} \Theta(t), \tag{S4}$$

and $\tau = \alpha_\mathrm{eff}^\phi \lambda / |\mathcal{F}'_\phi(\phi_\mathrm{eq})|$ is the relaxation time of $\phi$ degree of freedom. The correlation relation for the effective thermal fluctuation field $\tilde{\xi}_X(t)$ is given by the fluctuation-dissipation theorem $\langle \tilde{\xi}_X(t) \tilde{\xi}_X(t') \rangle \propto T f(|t-t'|)$ where $T$ is the temperature. The noise is 'colored' in the sense that it is no longer a white random noise.

### B. Order-of-magnitude estimation of $\tau$

To estimate the order of magnitude of $\tau$, we use the fact that the magnitude of $|\mathcal{F}_\phi|$ is determined by the DMI or the hard axis anisotropy: $|\mathcal{F}'_\phi(\phi_\mathrm{eq})| \approx \gamma \lambda (\pi/2) \times (2H_\perp \text{ or } D\lambda^{-1})$. We take the DMI field $D\lambda^{-1}$ being 30 mT [1] for a rough estimation. Then, $|\mathcal{F}'_\phi(\phi_\mathrm{eq})|/\lambda \approx \gamma \times 30$ mT $\approx 5$ GHz, so that $\tau = \alpha_\mathrm{eff}^\phi \lambda / |\mathcal{F}'_\phi(\phi_\mathrm{eq})| \approx \alpha_\mathrm{eff}^\phi \times 0.2$ ns, which is small compared to the time scale of the dynamics of $X$.

### C. First order approximation - chiral mass correction

Since $\tau$ is small, compared to the times scale of the dynamics of $X$, we may expand $f(t)$ by $\tau$, in the sense of the gradient expansion in time space. Then, $f(t) \approx \mathcal{L}[\alpha_{\text{eff}}^X + (1/\zeta_{\text{eff}}^2 \alpha_{\text{eff}}^\phi)s\tau] = \alpha_{\text{eff}}^X \delta(t) + (\tau/\zeta_{\text{eff}}^2 \alpha_{\text{eff}}^\phi)\delta'(t)$. Putting this into Eq. (S3) gives

$$\frac{\tau}{\zeta_{\text{eff}}^2 \alpha_{\text{eff}}^\phi} \frac{1}{\lambda} \frac{d^2 X}{dt} + \frac{\alpha_{\text{eff}}^X}{\lambda} \frac{dX}{dt} = \mathcal{F}_X + \tilde{\xi}_X(t), \tag{S5}$$

where the first term represents a massive term. To obtain the DW mass, we need to find the factor which makes $\mathcal{F}_X$ have the dimension of force. Note that the force generated by pushing the DW is calculated by $M_s \int \mathbf{H}_{\text{eff}} \cdot \partial_X \mathbf{m} d^3 x = (2M_s S/\gamma)\mathcal{F}_X$. Therefore, the mass is defined by multiplying the factor $2M_s S/\gamma$,

$$m_{\text{DW}} = \frac{1}{\zeta_{\text{eff}}^2} \frac{2M_s S\tau}{\gamma \alpha_{\text{eff}}^\phi \lambda}, \tag{S6}$$

which is equivalent to Eq. (5) in the main text.

### D. Higher order contributions - chiral creep

To calculate $v_0$, one needs to solve a barrier escape problem. For an energy barrier $E_b$, Kramer [2] derived the thermal escapes rate

$$\Gamma = \frac{\nu}{2\pi} \sqrt{\frac{|\mathcal{F}'(X_m)|}{|\mathcal{F}'(X_M)|}} e^{-E_b/k_B T}, \tag{S7}$$

where $\mathcal{F}'(X_m)$ and $\mathcal{F}'(X_M)$ are the derivatives of the force (second derivatives of the pinning energy landscape) evaluated at the potential well and the saddle point respectively. $\nu$ is called the reactive frequency [3] which we calculate below. Then, $v_0$ is proportional to $\Gamma$. According to the Kramer's theory, $\nu \propto 1/\alpha_{\text{eff}}^X$ for a high damping and Markovian limit, which was also confirmed by the functional renormalization group technique [4].

However, we generalize this result to a non-Markovian situation [Eq. (S3)]. To do this, we apply the theory of escape rate for a non-Markovian equation of motion [3, 5], based on which, the reactive frequency $\nu$ corresponding to Eq. (S3) is given by the positive root of the following algebraic equation:

$$\frac{1}{\lambda} \nu \mathcal{L}[f(t)](\nu) = |\mathcal{F}'(X_M)|, \tag{S8}$$

whose exact solution can be calculated from Eq. (S4). As a result,

$$\nu = \frac{2\nu_0}{(1-\tau\nu_0) + \sqrt{(1+\tau\nu_0)^2 + 4\tau\nu_0/\zeta_{\text{eff}}^2 \alpha_{\text{eff}}^X \alpha_{\text{eff}}^\phi}} \approx \begin{cases} \nu_0 \propto \dfrac{1}{\alpha_{\text{eff}}^X} & \nu_0\tau \ll \zeta_{\text{eff}}^2 \alpha_{\text{eff}}^X \alpha_{\text{eff}}^\phi, \\[2ex] \zeta_{\text{eff}} \sqrt{\dfrac{\nu_0 \alpha_{\text{eff}}^X \alpha_{\text{eff}}^\phi}{\tau}} \propto \zeta_{\text{eff}} & \nu_0\tau \gtrsim \zeta_{\text{eff}}^2 \alpha_{\text{eff}}^X \alpha_{\text{eff}}^\phi, \end{cases} \tag{S9}$$

where $\nu_0 = \lambda|\mathcal{F}'(X_M)|/\alpha_{\text{eff}}^X$ is the reactive frequency for $\tau = 0$. In the second limit, we assume that the damping parameters are small, thus the last term in the denominator in Eq. (S9) dominates the other terms in the denominator. The two limits shows completely different dependences of $\nu$ on the dynamical parameters. Therefore, it is important to determine the relevant regime.

Assuming $\mathcal{F}_{\text{pin}}$ is random, $|\mathcal{F}'(X_M)| \approx |\mathcal{F}'(X_m)|$ in Eq. (S7) gives $\nu_0/2\pi$ to be the typical attempt frequency $\approx 1$ GHz [6]. From $\tau \approx \alpha_{\text{eff}}^\phi \times 0.2$ ns estimated above, we obtain $\tau\nu_0 \approx \alpha_{\text{eff}}^\phi$ which is an order of magnitude larger than $\zeta_{\text{eff}}^2 \alpha_{\text{eff}}^X \alpha_{\text{eff}}^\phi$, thus the second regime in Eq. (S9) is more relevant, contrary to the traditional creep theory just taking $\tau = 0$.

---




\begin{thebibliography}{99}
\bibitem{footnote}
	To use the Onsager relation, one needs to assume a slow dynamics of magnetization, thus the coefficient in front of $\partial_t\vec{m}$ is quasistatic. Therefore, relativistic corrections~\cite{Mondal17PRB}, for example, cannot be considered in our theory.
\end{thebibliography}
\end{document}